# Universal condition for the existence of interface modes in the whole momentum space with arbitrary materials

J. W. Dong, J. Zeng, Q. F. Dai, and H. Z. Wang*

*State Key Laboratory of Optoelectronic Materials and Technologies,*

*Zhongshan (Sun Yat-Sen) University, Guangzhou 510275, China*

**Abstract**

It is shown that, by theoretical and experimental results, a universal zero-impedance condition exists for two kinds of localized interface modes in the whole momentum space (both above and below the light line). It can be applied at the interface between any two materials including photonic crystals, single-negative (negative-permittivity or negative-permeability) materials, or double-negative metamaterials. In addition, it presents an intuitive physical concept, and also provides a feasible way to determine interface modes, which will have predominance in various applications.

* Corresponding author, electronic address: stswhz@mail.sysu.edu.cn

## I. INTRODUCTION

An interface mode is a specific type of wave that is confined at the interface of a single-heterostructure (SHS) composed of two different media. It is always dependent on the interface properties, in particular on the material properties. Previous studies on interface mode focused on cases below the light line. For excitation of this kind of interface mode, an additional tangential wave vector should be introduced, which is analogous to electronic surface modes at solid boundaries, therefore many interface modes (below the light line) are also called surface modes.[1] Recently, an interface mode was found not only in simple homogenous materials, but also in complex systems like a periodic array of dielectric materials, i.e. photonic crystals (PCs).[2,3] In addition, interface modes have also extended to the metamaterials[4-7] which have been developed rapidly these years.

Meanwhile, another type of localized interface mode, which is present above the light line and excited in any tangential wave vector including zero wave vector, has also been reported. Until now, it has been theoretically studied at the interface of some complex SHS, e.g. PC-PC heterostructures[8,9] that two PCs with different lattices and different physical and geometrical parameters are connected directly. The remarkable characteristics of both kinds of interface modes will lead to many applications, such as polariton lasers, thermal emission, directional emission, monolayer research, and so on.[10-13]

In fact, the physical inbeing of both interface modes is the result of mode matching between structures on two sides of the interface. Although the interface mode above the light line can be understood by phase matching in a "zero-thick" cavity, there is lack of an intuitive physical concept to understand both interface modes simultaneously. Therefore, it is in need to obtain a universal condition that governs the existence of these interface modes for arbitrary SHS composed of any two materials in the whole momentum space, i.e. both above and below the light line. This condition will provide a simple and feasible way to reveal the interface mode no matter where it is, while increasing its use in future applications.

In this paper, impedance spectra of different materials have been investigated. Zero impedance (zero-$Z$) condition is proposed for the presence of an interface mode in the whole momentum space. This condition is universal for arbitrary SHS composed of any two simple or complex materials.

## II. THEROICAL MODEL: ZERO IMPEDANCE CONDITION

Consider an electromagnetic wave incident upon a general 1D SHS composed of any two materials (a piece of homogenous slab or complex stacks). The ratio of total intensity $E_{tot}$ to incident intensity $E_{in}$ at the interface of left and right materials can be obtained by self-consistent conditions:

$$\left|\frac{E_{tot}}{E_{in}}\right|^2 = \frac{\left[1-\left(r^l\right)^2\right]\left[\left(r^r-1\right)^2 + 4r^r \sin^2\left(\arctan \chi^r\right)\right]}{\left(r^l - r^r\right)^2 + 4r^l r^r \sin^2\left(\arctan \dfrac{\chi^l + \chi^r}{1 - \chi^l \chi^r}\right)}$$

where $r^{l,r}$ and $\chi^{l,r}$ are the reflection modulus and the imaginary part of the effective impedance ($Z_{eff}^{l,r}$: purely imaginary number in forbidden regions, i.e. $Z_{eff}^{l,r} = i\chi^{l,r}$) of the separate left and right materials, respectively. When the zero-$Z$ condition occurs ($\chi^l + \chi^r = 0$), the denominator in right side of above equation become minimum, and the intensity ratio at the interface between the left and right materials will increase rapidly. If the electromagnetic wave were evanescent in the two materials respectively, the interface mode would be present. For normal incidence, the effective impedance of a piece of homogenous slab is equivalent to material's intrinsic impedance. But this is not the case for oblique incidence and for complex systems. Instead, it should be obtained by either reflection phase-shift or wave impedance at the surface that is defined by the ratio of transverse fields ( $Z_{surf}^{l,r} = j\chi_{surf}^{l,r} = E_{//}^{l,r} / H_{//}^{l,r}$ ). For usual excitation (above the light line), two kinds of impedances by using different methods are equivalent. In this case, zero-$Z$ condition can be regarded as the phase matching condition with zero cavity thickness. However, for unusual excitation (below the light line), the phase matching condition is no use since the phase-shift is always zero. In this case, wave impedance should be employed to find out the interface mode. It is noticed that the left and right materials are not specified yet. That is to say the zero-$Z$ condition is a universal condition for the existence of an interface mode in arbitrary SHS composed of any two materials, e.g. PCs, single-negative (SNG) materials, and double-negative (DNG) metamaterials.

The schematic impedance spectra of two types of PCs are shown in Fig. 1(a) and (b). Here, HL-PC (LH-PC) means the first PC layer from the interface is the high (low) refractive index layer. The impedance spectra of negative-permittivity (ENG) and negative-permeability (MNG) materials are also shown in Fig. 1(c) and (d). It is found that the zero-$Z$ condition can be satisfied in various frequency regions. For example, it is in region 1 and 3 for PC-PC configuration, as well as in region 2 for ENG and HL-PC configuration and in region 1 for MNG and LH-PC configuration, and so on. That is to say the interface modes are present in these frequencies.

## III. EXPERIMENENTAL DEMOSTRATIONS AND NUMERICAL SIMULATIONS

### A. ENG-PC configuration

Let us first consider a particular SHS configuration: ENG and HL-PC configuration [shown in Fig. 2(a)], consisting of a silver film and PC with their structural parameters (shown in the caption of Fig. 3) that ensure the zero-$Z$ condition is satisfied. The impedance spectra of two sub-structures (the silver film[14] and the HL-PC) and the whole SHS are shown in Fig. 3(a). Two zero-Z points, $f_1 = 0.2180$ (1.42 μm) and $f_2 = 0.4708$ (0.658 μm), are present in the forbidden region of two separate sub-structures. These points are demonstrated by both the calculations and experiments. Although the transmittance of 50nm-thick silver film and HL-PC are opaque in the studied frequency range, $f_1$ experiences high transmittance in the whole SHS [without thickness error, dot line in Fig. 3(b)]. This is the same to another zero-$Z$ point $f_2$. Meanwhile, in the experiment, the 1D sample was fabricated using ZnS ($n$=2.35) and MgF$_2$ ($n$=1.38) by the vacuum coating technique. The measured transmission spectrum [solid line in Fig. 3(b)] was performed on a spectrophotometer (Perkin-Elmer, model Lambda-900). It exhibits a resonant transmission peak at 0.2315 (1.34 μm) which is consistent of the zero-$Z$ point $f_1$ [dash line in Fig. 3(b)] with thickness errors in each layer.[15] In addition, we have also performed many other experiments and calculations. The results show that the interface mode would be obtained if the layer thicknesses were controlled precisely.

Figure 3(c) shows the intensity distributions whether the zero-$Z$ condition is present or not. One can see that when the condition is present (e.g. $f_1$), the intensity is significantly enhanced in the layer next to ENG material (metal here). Moreover, the intensity decays away from the interface exponentially. Contrastingly, when the condition is absent (e.g. $f_3$), the intensity decreases in the metallic layer, but oscillates in the PC because $f_3$ locates the propagating frequency band of PC.

**B. PC-PC configuration**

Next, we consider a PC-PC heterostructure [shown in Fig. 2(b)]. By calculating the effective impedances, we find out a set of structural parameters (shown in the caption of Fig. 4) to meet zero-$Z$ condition. Figure 4(a) shows that the transmission peaks extracted from the measured spectra are from 0.4928 (0.838 μm) to 0.6155 (0.671 μm). In order to demonstrate the intuitive concept on the zero-$Z$ condition, the rigorous formulas in a framework of general transfer matrix method with arbitrary frequency-dependent materials are employed, and the complex transcendent equation to estimate the interface mode can be expressed as:

$$\frac{P^l(1,2)}{\exp(iK_B^l\Lambda^l)-P^l(1,1)} - \frac{Q(1,1)P^r(1,2)+Q(1,2)\left[\exp(-iK_B^r\Lambda^r)-P^r(1,1)\right]}{Q(2,1)P^r(1,2)+Q(2,2)\left[\exp(-iK_B^r\Lambda^r)-P^r(1,1)\right]} = 0$$

The derivation of the above equation is shown in *Appendix* in detail. The solutions are also shown in Fig. 4(a). It is found that the experiment result is well consistent of the rigorous theory. Figure 4(b) shows the intensity distribution of the interface mode at $f_4 = 0.4928$ (0.838 μm) for normal incidence. The profile of localization and exponential decay behavior are similar to the ENG-PC configuration.

**C. All-metamaterials-based configuration**

In the above analysis, we focus on the existence of interface modes above the light line. In the following, we extend it to below the light line. We choose the configuration of DNG metamaterials and PC stacking with SNG materials. The structural parameters are shown in the Fig. 5 caption. Projected band structure and the wave impedance spectra in different tangential wavevector are being studied. The zero-impedance contour line is shown in Fig. 5(a). It is found that the contour line locates in the decay regions of both DNG metamaterials and PC stacking with SNG materials. The unique properties of the interface mode at $f_5 = 5$ GHZ when $k_x = 2(2\pi/\Lambda^r)$ are also shown in Fig. 5(b). It is interesting to find that the intensity decays more rapidly than in the cases above. This means the degree of localization is much stronger in the all-metamaterials-based SHS than the former cases. In addition, we also find the interface mode below the light line in PC-PC heterostructures stacking with dielectric when the wave impedances are compensated.

## IV. CONCLUSIONS

In conclusion, a universal zero-$Z$ condition has been proposed in the whole momentum space for the existence of an interface mode in arbitrary single-heterostructures, with any configuration of materials. The experimental and theoretical results on PC-PC, as well as ENG-PC and DNG-PC, are employed to clarify this idea in optical and in gigahertz, respectively. Although only 1D case has been demonstrated, similar rules will appear in multidimensional single-heterostructures.[16] Moreover, our remarkable concept provides a general method to find interface modes in various applications.

**ACKNOWLEDGMENTS**

The authors gratefully acknowledge Prof. C. T. Chan (Hong Kong University of Science and Technology) for his valuable comments. This work is supported by the National Natural Science Foundation of China (10674183) and National 973 Project of China (2004CB719804).

**APPENDIX**

Consider a two-component 1D PC structure with arbitrary dispersive materials [c.f. Fig. 2(b)]. The structure can be described by three structural parameters: $\varepsilon_j, \mu_j, d_j$, where $\varepsilon_j, \mu_j$ and $d_j$ are the permittivity, permeability and thickness of the $j$th layer, $j = 1,2$. Then the electric field distribution in the $j$th layer can be expressed as:

$$E_{j,y}(x,z) = \left[ a_j \exp(-ik_{jz}d_j) + b_j \exp(ik_{jz}d_j) \right] \exp(ik_x x) \tag{A1}$$

where $k_{jz} = \left[ \varepsilon_j \mu_j (\omega/c)^2 - \beta^2 \right]^{1/2}$, and $\beta$ is the wavevector component in x direction, $a_j$ and $b_j$ are the amplitudes at the interface for the ''forward'' and ''backward'' waves, respectively. $\omega$ is the angular frequency the incidence electromagnetic wave, and $c$ is the velocity of light in the vacuum. Since the interface modes are lying in the forbidden gaps of the left and right PC, the electromagnetic wave becomes evanescent. The amplitudes of the ''forward'' wave decays exponentially as $z$ increases and the amplitude of the ''backward'' wave grows exponentially with $z$. Thereby, the Bloch wave number $K_B$ should take the complex value in the form as

$$K_B^{l,r} = \frac{m^{l,r}\pi}{\Lambda^{l,r}} + iq^{l,r} \quad \left(q^{l,r} > 0, m^{l,r} = 0,1,2,...\right) \tag{A2}$$

where $\Lambda$ is the period of unit cell of PC, and the superscript $l$ and $r$ represent the left and right PC, respectively. The coefficients of each PC are related by

$$\left(I - P^l \exp(-iK_B^l \Lambda^l)\right)\begin{pmatrix} a_2^l \\ b_2^l \end{pmatrix} = 0 \tag{A3}$$

for the left PC, and

$$\left(I - P^r \exp(iK_B^r \Lambda^r)\right)\begin{pmatrix} a_1^r \\ b_1^r \end{pmatrix} = 0 \tag{A4}$$

for the right PC, respectively. The coefficients of adjacent layers between each PC are related by

$$\begin{pmatrix} a_2^l \\ b_2^l \end{pmatrix} = Q \begin{pmatrix} a_1^r \\ b_1^r \end{pmatrix} \tag{A5}$$

The elements of matrix $P^{l,r}$ are

$$P^l(1,1) = \zeta_{12}^l - \eta_{12,+}^l \xi_{12}^l + i\left(\psi_{21}^l + \eta_{12,+}^l \psi_{12}^l\right), \quad P^l(1,2) = \eta_{12,-}^l \left(\xi_{12}^l + i\psi_{12}^l\right),$$

$$P^l(2,1) = \eta_{12,-}^l \left(\xi_{12}^l - i\psi_{12}^l\right), \quad P^l(2,2) = \zeta_{12}^l - \eta_{12,+}^l \xi_{12}^l + i\left(\psi_{21}^l - \eta_{12,+}^l \psi_{12}^l\right), \tag{A6}$$

$$P^r(1,1) = \zeta_{21}^r - \eta_{21,+}^r \xi_{21}^r + i\left(\psi_{12}^r + \eta_{21,+}^r \psi_{21}^r\right), \quad P^r(1,2) = \eta_{21,-}^r \left(\xi_{21}^r + i\psi_{21}^r\right),$$

$$P^r(2,1) = \eta_{21,-}^r \left(\xi_{21}^r - i\psi_{21}^r\right), \quad P^r(2,2) = \zeta_{21}^r - \eta_{21,+}^r \xi_{21}^r + i\left(\psi_{12}^r - \eta_{21,+}^r \psi_{21}^r\right), \tag{A7}$$

where $\zeta_{12}^{l,r} = \cos k_{1z}^{l,r} d_1^{l,r} \cos k_{2z}^{l,r} d_2^{l,r}$, $\xi_{12}^{l,r} = \sin k_{1z}^{l,r} d_1^{l,r} \sin k_{2z}^{l,r} d_2^{l,r}$, $\psi_{12}^{l,r} = \sin k_{1z}^{l,r} d_1^{l,r} \cos k_{2z}^{l,r} d_2^{l,r}$, and $\left(\eta_{12,\pm}^{l,r}\right)^{TE} = \frac{1}{2}\left(\mu_2^{l,r} k_{1z}^{l,r} / \mu_1^{l,r} k_{2z}^{l,r} \pm \mu_1^{l,r} k_{2z}^{l,r} / \mu_2^{l,r} k_{1z}^{l,r}\right)$ for TE wave, $\left(\eta_{12,\pm}^{l,r}\right)^{TM} = \frac{1}{2}\left(\varepsilon_1^{l,r} k_{2z}^{l,r} / \varepsilon_2^{l,r} k_{1z}^{l,r} \pm \varepsilon_2^{l,r} k_{1z}^{l,r} / \varepsilon_1^{l,r} k_{2z}^{l,r}\right)$ for TM wave, respectively. The elements of matrix $Q$ are

$$Q^{TE}(1,1) = \frac{1}{2}\left(1 + \mu_2^l k_{1z}^r / \mu_1^r k_{2z}^l\right)\exp(ik_{1z}^r d_1^r), \quad Q^{TE}(1,2) = \frac{1}{2}\left(1 - \mu_2^l k_{1z}^r / \mu_1^r k_{2z}^l\right)\exp(-ik_{1z}^r d_1^r),$$

$$Q^{TE}(2,1) = \frac{1}{2}\left(1 - \mu_2^l k_{1z}^r / \mu_1^r k_{2z}^l\right)\exp(ik_{1z}^r d_1^r), \quad Q^{TE}(2,2) = \frac{1}{2}\left(1 + \mu_2^l k_{1z}^r / \mu_1^r k_{2z}^l\right)\exp(-ik_{1z}^r d_1^r),$$

$$\tag{A8}$$

for TE wave, and

$$Q^{TM}(1,1) = \frac{1}{2}\left(\sqrt{\varepsilon_2^l}\sqrt{\mu_2^l}/\sqrt{\varepsilon_1^r}\sqrt{\mu_1^r}\right)\left(k_{1z}^r/k_{2z}^l\right)\left(1+\varepsilon_1^r k_{2z}^l/\varepsilon_2^l k_{1z}^r\right)\exp\left(ik_{1z}^r d_1^r\right),$$

$$Q^{TM}(1,2) = \frac{1}{2}\left(\sqrt{\varepsilon_2^l}\sqrt{\mu_2^l}/\sqrt{\varepsilon_1^r}\sqrt{\mu_1^r}\right)\left(k_{1z}^r/k_{2z}^l\right)\left(1-\varepsilon_1^r k_{2z}^l/\varepsilon_2^l k_{1z}^r\right)\exp\left(-ik_{1z}^r d_1^r\right),$$

$$Q^{TM}(2,1) = \frac{1}{2}\left(\sqrt{\varepsilon_2^l}\sqrt{\mu_2^l}/\sqrt{\varepsilon_1^r}\sqrt{\mu_1^r}\right)\left(k_{1z}^r/k_{2z}^l\right)\left(1-\varepsilon_1^r k_{2z}^l/\varepsilon_2^l k_{1z}^r\right)\exp\left(ik_{1z}^r d_1^r\right),$$

$$Q^{TM}(2,2) = \frac{1}{2}\left(\sqrt{\varepsilon_2^l}\sqrt{\mu_2^l}/\sqrt{\varepsilon_1^r}\sqrt{\mu_1^r}\right)\left(k_{1z}^r/k_{2z}^l\right)\left(1+\varepsilon_1^r k_{2z}^l/\varepsilon_2^l k_{1z}^r\right)\exp\left(-ik_{1z}^r d_1^r\right), \quad (A9)$$

for TM wave, respectively. In addition, from the eigenequation Eqs. (A3) and (A4), the eigenvalues are:

$$\exp\left(iK_B^{l,r}\Lambda^{l,r}\right) = \frac{1}{2}\left[P^{l,r}(1,1)+P^{l,r}(2,2)\right] \pm \left\{\frac{1}{4}\left[P^{l,r}(1,1)+P^{l,r}(2,2)\right]^2 - 1\right\}^{1/2} \quad (A10)$$

The choice of the eigenvalues is to ensure $\exp\left(iK_B^{l,r}\Lambda^{l,r}\right) < 1$ for evanescent wave. The eigenvector is

$$\begin{pmatrix} a_2^l \\ b_2^l \end{pmatrix} = \begin{bmatrix} P^l(1,2) \\ \exp\left(iK_B^l\Lambda^l\right) - P^l(1,1) \end{bmatrix} \quad (A11)$$

for the left PC, and

$$\begin{pmatrix} a_1^r \\ b_1^r \end{pmatrix} = \begin{bmatrix} P^r(1,2) \\ \exp\left(-iK_B^r\Lambda^r\right) - P^r(1,1) \end{bmatrix} \quad (A12)$$

for the right PC, respectively. Finally, employing Eqs. (A5), (A11), and (A12), the complex transcendent equation to estimate the interface mode can be expressed as:

$$\frac{P^l(1,2)}{\exp\left(iK_B^l\Lambda^l\right) - P^l(1,1)} - \frac{Q(1,1)P^r(1,2) + Q(1,2)\left[\exp\left(-iK_B^r\Lambda^r\right) - P^r(1,1)\right]}{Q(2,1)P^r(1,2) + Q(2,2)\left[\exp\left(-iK_B^r\Lambda^r\right) - P^r(1,1)\right]} = 0 \quad (A13)$$

For a given $\omega$ and $\beta$, the terms in Eq. (A13) can be obtained from Eqs. (A6)-(A10) so that it can be solved numerically.

**Figure Captions.**

FIG. 1. (Color online) Impedance spectra of (a) HL-PC, (b) LH-PC, (c) ENG, and (d) MNG materials in the forbidden region. Here, HL-PC (LH-PC) means the first PC layer from the interface is the high (low) refractive index layer. The Drude model of $\varepsilon$ and $\mu$ is used here.

FIG. 2. (Color online) Schematic diagrams of SHS configurations: (a) metamaterial-PC and (b) PC-PC. $\varepsilon_i^{l,r}, \mu_i^{l,r}, d_i^{l,r}$ and $\Lambda^{l,r}$ are permittivity, permeability, thickness, and lattice constants of left (right) sub-structures, respectively.

FIG.3. (Color online) (a) Impedance spectra of SHS (solid), silver film (dot), and PC (dash). The grey/white denotes the propagating/evanescent mode of PC. (b) Measured (solid) and calculated with/without error (dash/dot) transmission spectra. (c) Intensity distribution with/without (solid/dot) localization. The structural parameters are $n_H^r = 2.35, n_L^r = 1.38, d_H^r = 0.198 \mu m, d_L^r = 0.112 \mu m, N^r = 6, d_{Ag} = 0.05 \mu m$.

FIG. 4. (Color online) (a) Projected band structure of PC-PC SHS with light lines (red lines), showing an interface mode (theory: blue lines; measure: open circles). Propagating modes of both PCs (black); evanescent modes of either PC (white). (b) Intensity distribution of the interface mode. The structural parameters are $n_H^r = 2.35, n_L^r = 1.38, d_H^l = 0.28 \mu m, d_L^l = 0.38 \mu m, d_H^r = 0.223 \mu m, d_L^r = 0.19 \mu m, N^r = 6$.

FIG. 5. (Color online) (a) Projected band structure of all-metamaterials-based SHS, showing interface modes (black line). Propagating modes of DNG (green) and PC (blue); evanescent modes for either sub-structures (white). (b) Intensity distribution of the interface mode at $f = 5\text{GHz}, k_x = 2(2\pi/\Lambda^r)$. The material's dispersions are similar to those of Ref. 4 and 5. The thicknesses are $d_{DNG} = 10mm, d_{N\varepsilon} = 8mm, d_{N\mu} = 16mm$.

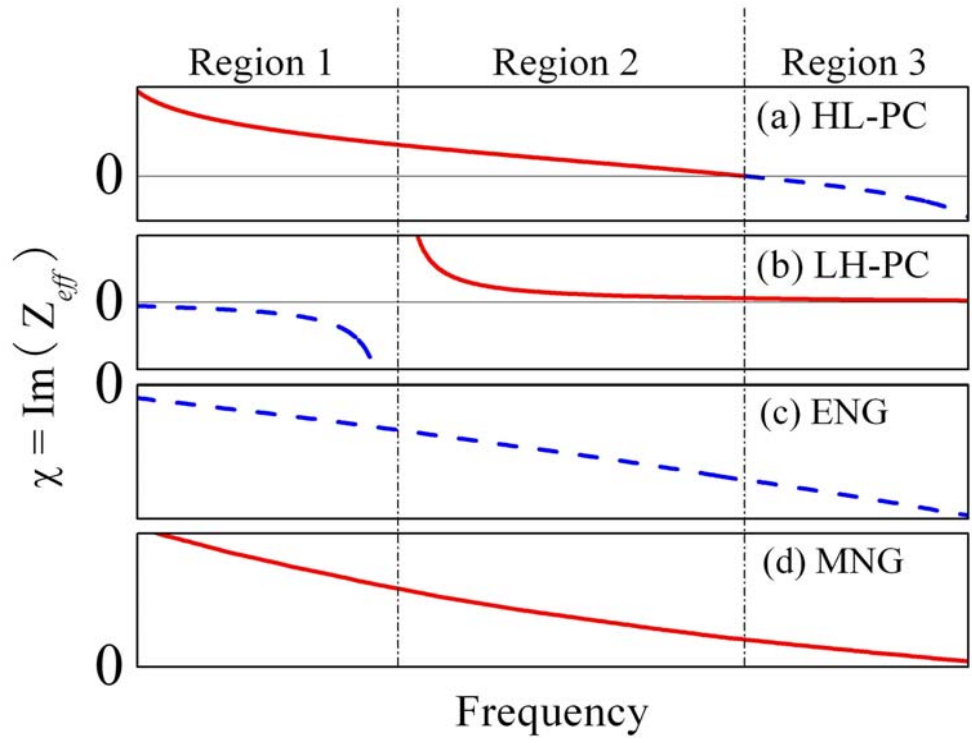

FIG. 1. Dong *et.al.*

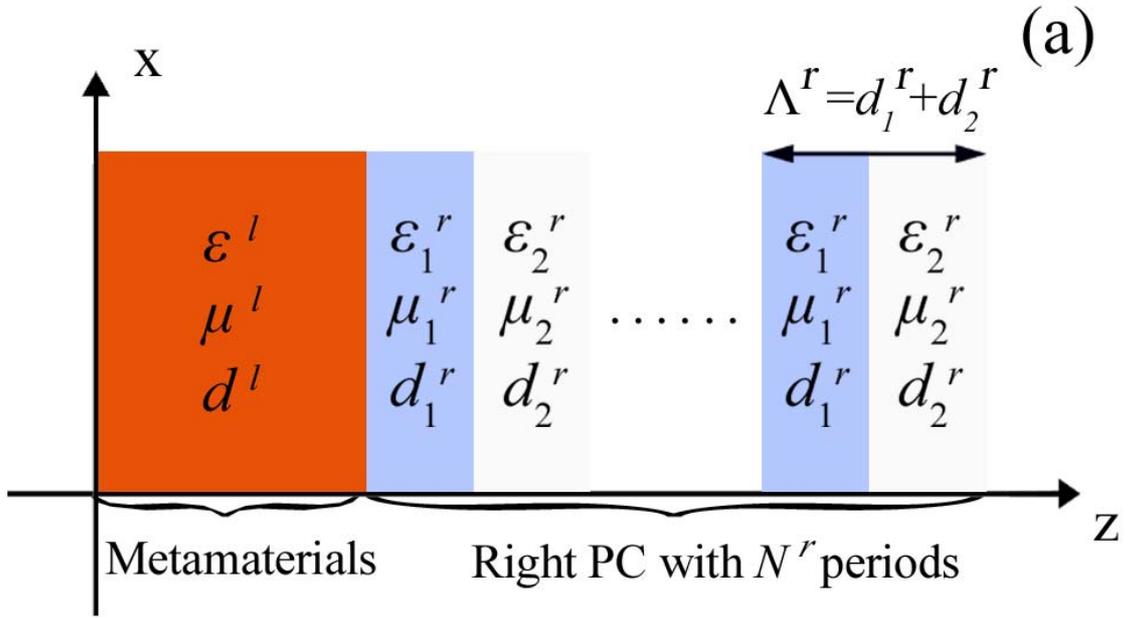

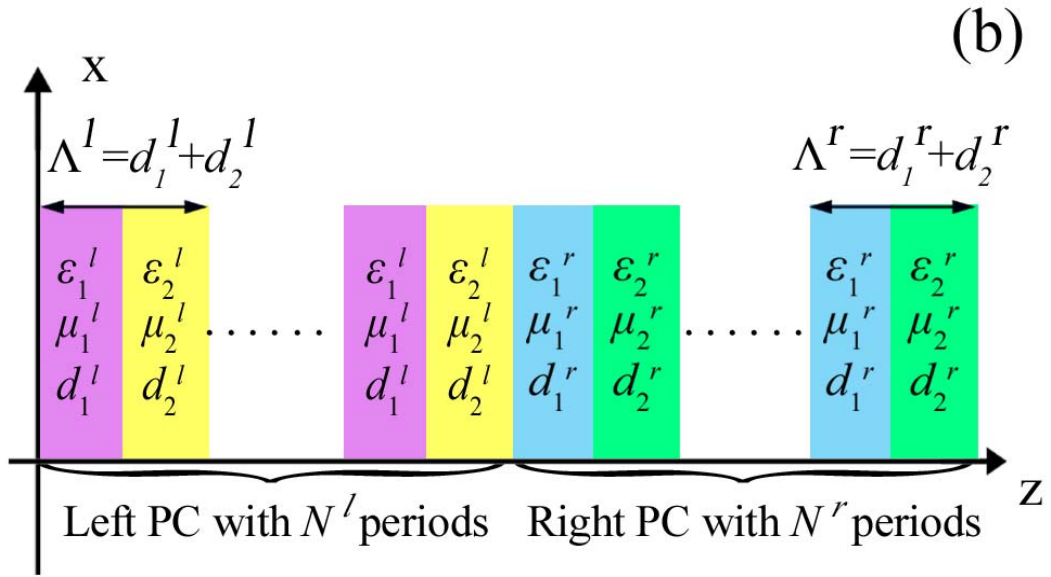

FIG. 2. Dong *et.al.*

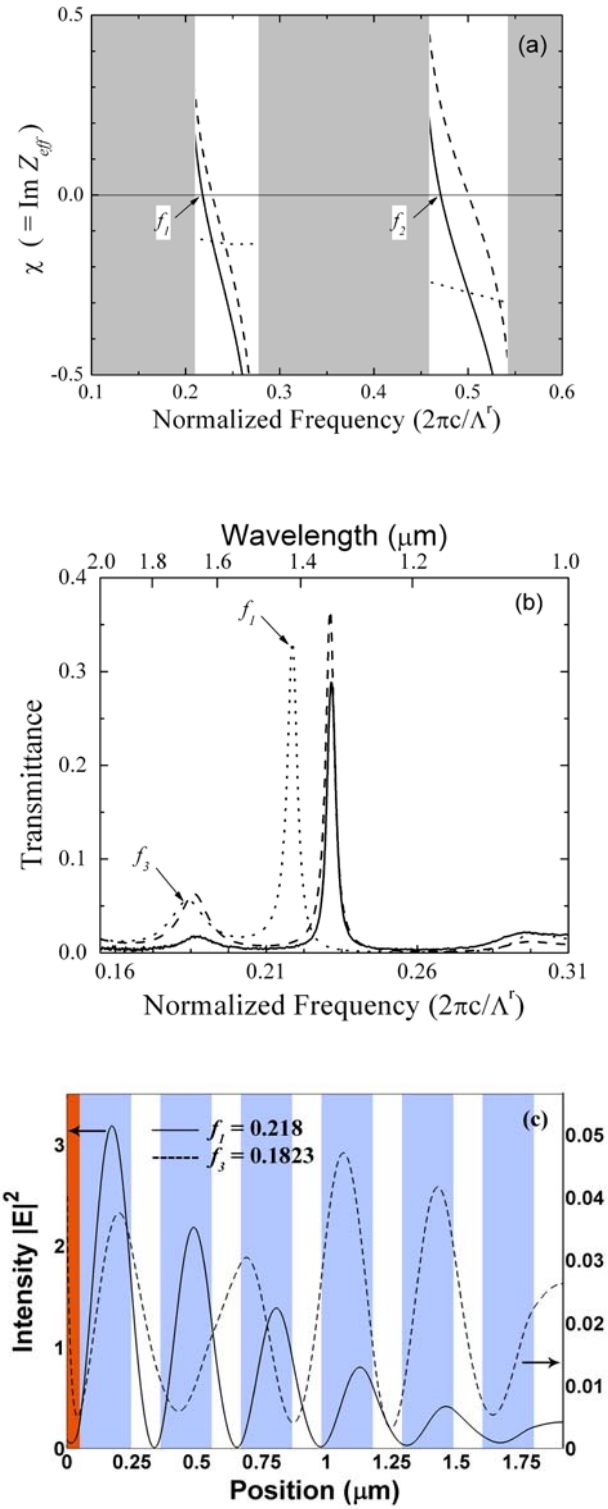

FIG. 3. Dong *et.al.*

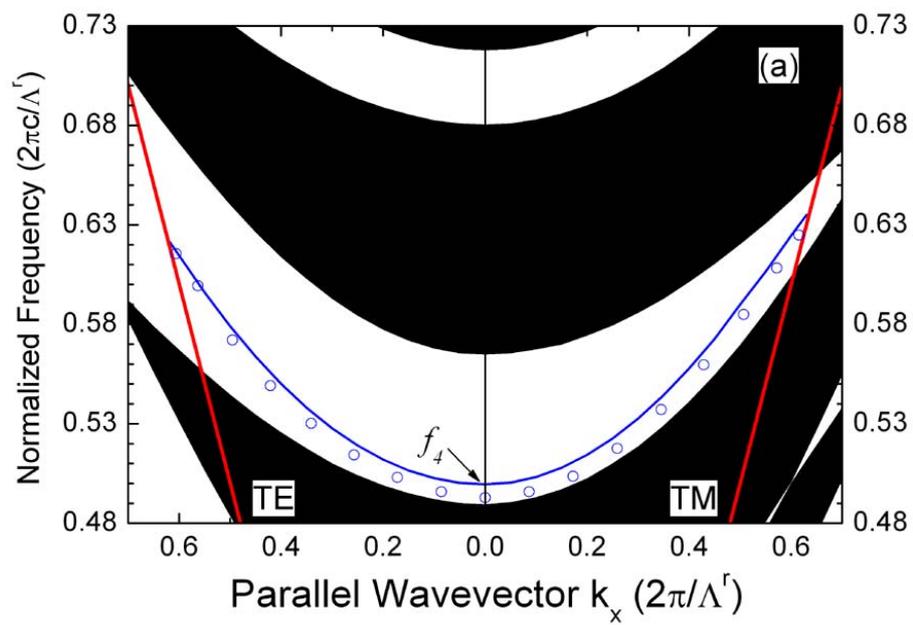

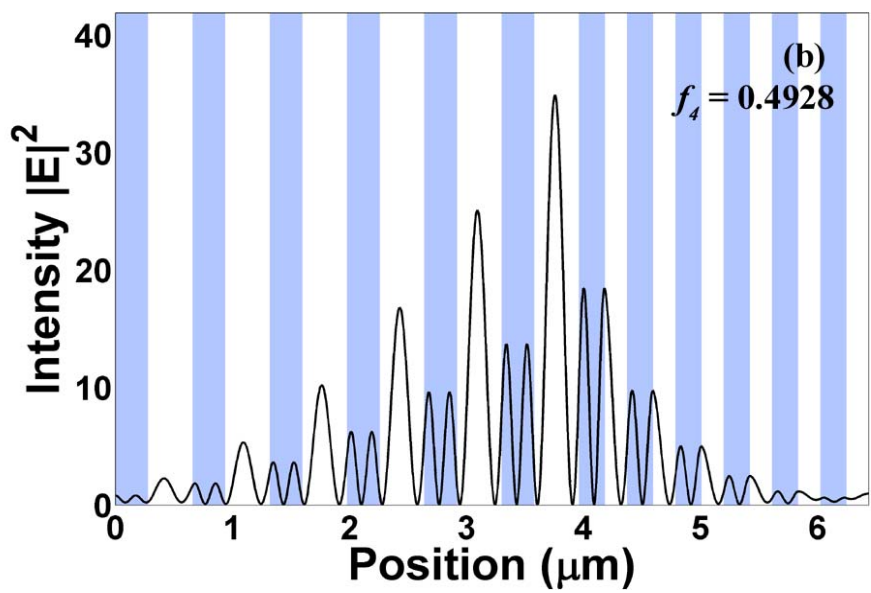

FIG. 4. Dong *et.al.*

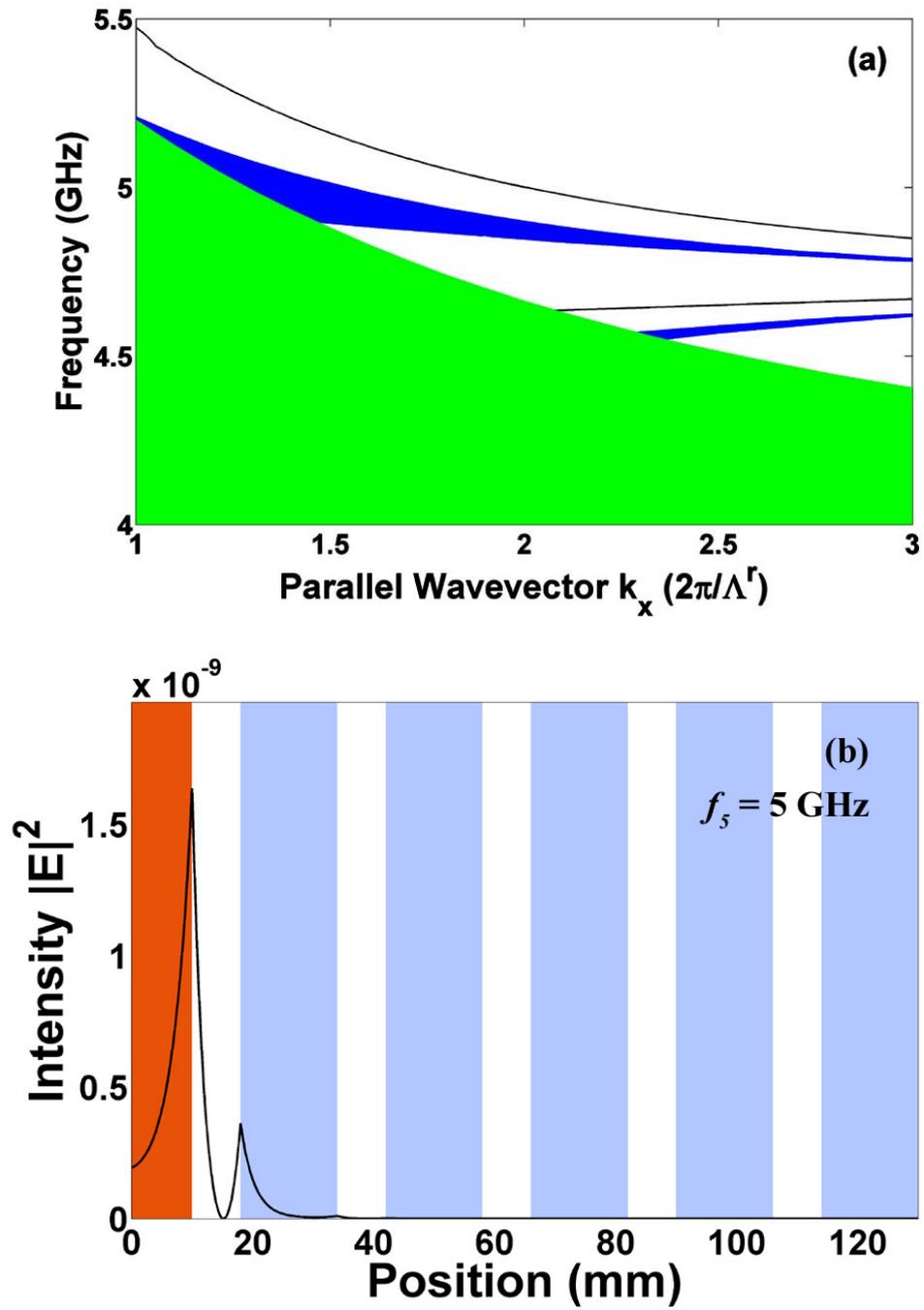

FIG. 5. Dong *et.al.*